\documentclass[fleqn,usenatbib]{mnras}

\usepackage{newtxtext,newtxmath}

\usepackage[T1]{fontenc}

\DeclareRobustCommand{\VAN}[3]{#2}
\let\VANthebibliography\thebibliography
\def\thebibliography{\DeclareRobustCommand{\VAN}[3]{##3}\VANthebibliography}

\usepackage[pdftex]{graphicx}
\usepackage{enumitem}
\usepackage{mathtools}
\usepackage[version=4]{mhchem}

\DeclarePairedDelimiterX\braket[2]{\langle}{\rangle}{#1 \delimsize\vert #2}
\usepackage{float}
\usepackage{array,multirow,graphicx}
\usepackage[dvipsnames]{xcolor}

\title[Global 21-cm: Ionospheric Effects]{Quantifying Ionospheric Effects on Global 21-cm Observations}

\author[Emma Shen et al.]{
Emma Shen,$^{1}$\thanks{E-mail: yhs24@cam.ac.uk}
Dominic Anstey,$^{1}$
Eloy de Lera Acedo,$^{1,2}$
Anastasia Fialkov,$^{2,3}$
Will Handley$^{1,2}$
\\
$^{1}$Cavendish Laboratory, University of Cambridge, Cambridge, CB3 0HE, United Kingdom\\
$^{2}$Kavli Institute for Cosmology, Madingley Road, Cambridge, CB3 0HA, United Kingdom\\
$^{3}$Institute of Astronomy, University of Cambridge, Madingley Road, Cambridge CB3 0HA, United Kingdom
}

\date{Accepted XXX. Received YYY; in original form ZZZ}

\pubyear{2015}

\begin{document}
\label{firstpage}
\pagerange{\pageref{firstpage}--\pageref{lastpage}}
\maketitle

\begin{abstract}
 We modelled the two major layer of Earth's ionosphere, the F-layer and the D-layer, by a simplified spatial model with temporal variance to study the chromatic ionospheric effects on global 21-cm observations. From the analyses, we found that the magnitude of the ionospheric disruptions due to ionospheric refraction and absorption can be greater than the expected global 21-cm signal, and the variation of its magnitude can differ, depending on the ionospheric conditions. Within the parameter space adopted in the model, the shape of the global 21-cm signal is distorted after propagating through the ionosphere, while its amplitude is weakened. It is observed that the ionospheric effects do not cancel out over time, and thus should be accounted for in the foreground calibration at each timestep to account for the chromaticity introduced by the ionosphere.  

\end{abstract}




\section{Introduction}

A direct measurement of neutral hydrogen (HI) by measuring the 21-cm hyperfine transition line can provide a way to trace the intergalactic medium over time. However, the strong galactic foreground compared with the extremely weak signal is one of the main limitations of current 21-cm experiments, and in some cases their effects can be up to five orders of magnitude greater than the cosmological signal. In 2018, \citet{edges} reported to have found an absorption profile in the form of a \textit{flattened} Gaussian centred at 78 MHz in the integrated spectrum by fitting it with the foreground model and a model for the 21-cm signal simultaneously. However, concerns have been raised for the unphysical parameters implied by their data, and the non-uniqueness of their solution \citep{hillsnat, singh, bevins, sims}.

The strong galactic foreground compared with the extremely weak signal is one of the main difficulties in achieving a detection of the 21-cm signal. Other astronomical radio sources emitting in the frequency band 30-240 MHz together are $\sim$ 4-5 orders of magnitude larger than the 21-cm signal in brightness. These foreground sources include galactic synchrotron radiation, thermal free-free emission from galactic dust, and emissions from extragalactic point sources. Thus, precise understanding of frequency-dependent or chromatic effects of instrumental gain, instrumental noise contribution, antenna beam shape and ionospheric effects, coupled with spatially and spectrally varying foregrounds is critical to measuring the global 21-cm signal. 
 
 These effects are particularly challenging for detecting the global 21-cm signal because its relevant band lies in the lower frequency range; galactic foreground brightness temperature, with galactic synchrotron radiation being the major constituent, increases with decreasing frequency as a power law $T_\mathrm{sync} \propto \nu^{-\alpha}$ with a spectral index of $\alpha \sim -2.54$ \citep{edges}. These foregrounds are considered spectrally \textit{smooth}. It seems plausible that the signal could be extracted simply by fitting a power law. However, the inevitable chromaticity introduced by the antenna negates this assumption. Moreover, the increased fractional bandwidth in the lower frequency band leads to an increased variation of antenna beams across the measurement bandwidth, resulting in larger chromatic effects. Neither chromatic antenna beam nor the sky temperature is smooth as a function of frequency; after integration, the convolution would break the assumed smoothness.

The ionosphere further complicates chromatic mixing. The ionosphere is a magnetised plasma situated at the upper atmosphere spanning from $\sim$ 50-600 km above the Earth’s surface. This ionised layer can corrupt a propagating signal. The most significant propagation effects relevant to the sky-averaged spectra are absorption and refraction. These effects increase with decreasing frequency, which scale approximately as $\nu^{-2}$, and are 2-3 orders of magnitude larger than the global 21-cm signal \citep{km, burns}. This poses an especially large problem to experiments that adopt a wide field of view. In a global experiment, in order to achieve an achromatic antenna beam, it is necessary to adopt a wide field of view, as narrow beams often come with the penalty of chromatic effects. Also, a narrow-beam global experiment would be subject to cosmic variance \citep{cos}. The global 21-cm signal could be measured directly by subtracting the modelled foreground contribution from the antenna temperature were our prior knowledge on the foregrounds and chromatic mixing accurate enough, without which we are left to make simple assumptions on those properties. 

This work aims to address the chromatic effects due to the signal propagation through the temporally varying ionosphere, to which a satisfying solution has yet to be proposed. It analyses the chromatic ionospheric effects on global 21-cm observations by simulating the two major ionospheric layers, the F-layer and the D-layer, by a simplified spatial model with temporal variance. The analysis focuses on the chromatic corruptions introduced by the ionosphere after applying a chromaticity correction factor. The model incorporates actual ionospheric data during nighttime, the time of day favourable to observations for Radio Experiment for the Analysis of Cosmic Hydrogen (REACH) \citep{reachh}.

\section{Ionospheric Model}

The model adopted in this work is based on \citet{km}; in their paper, it is shown that chromatic refraction and absorption on incoming rays due to the ionosphere would introduce a significant distortion to the foregrounds. Likewise, the model adopted in this paper considers only ionospheric refraction and absorption. Dynamic effects such as scintillation induced by ionospheric turbulence \citep{crane}, refraction resulted from large scale traveling ion-disturbances \citep{boug1991}, etc. are assumed to cancel out over time. Effects due to magnetic fields are smaller than day to day changes in electron density and thus are also ignored. Furthermore, Faraday rotation is assumed to be unpolarised on scales comparable to the antenna beam \citep{km}. The ionospheric layers are assumed to be homogeneous, and have no spatial variation across the azimuth.

\begin{figure}
    \centering
        \minipage{0.532\textwidth}
        \includegraphics[trim={0.21cm 0.7cm 0 6.cm},clip,width=\linewidth]{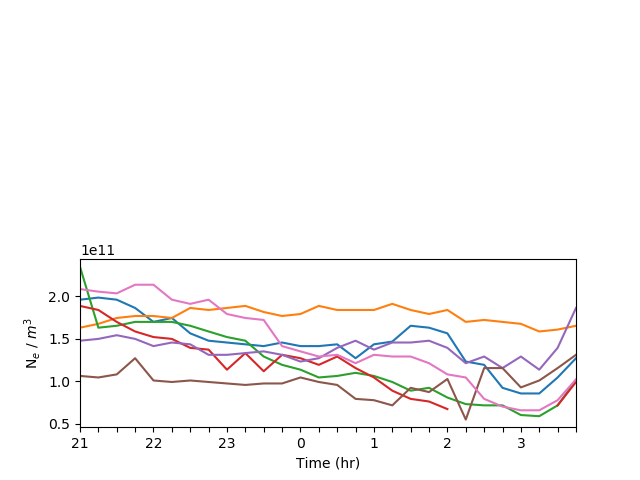}
        \endminipage\hfill
        \minipage{0.532\textwidth}
        \includegraphics[trim={0.21cm 0 0 6.48cm},clip,width=\linewidth]{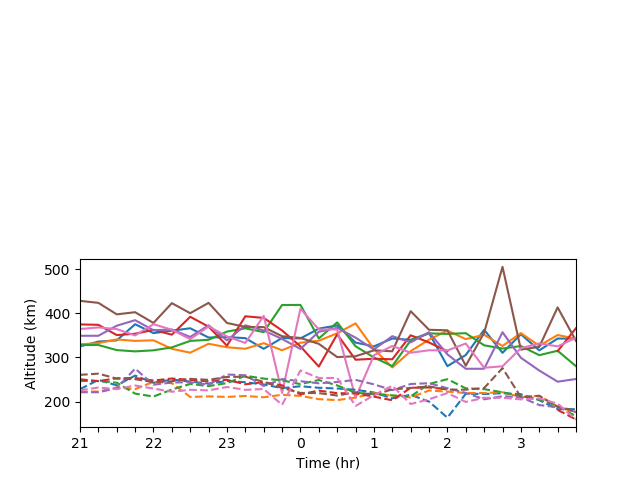}
        \endminipage\hfill
    \caption{Ionospheric data retrieved from Lowell GIRO Data Center at station Louisvale, South Africa. Each colour represents a different night. The upper panel shows the maximum electron density in the F-layer overnight. The lower panel shows the upper F-layer boundary in solid lines and the lower F-layer boundary in dashed lines.}
    \label{figdata}
\end{figure}

In this paper, we simulate the ionosphere with time-varying parameters using actual data for the F-layer (Fig. \ref{figdata}).  The data is collected from Lowell GIRO Data Center at station Louisvale, South Africa (\url{https://ulcar.uml.edu/DIDBase/}), whose geological location is close to where REACH is situated (the Karoo, South Africa). There is currently no available data required to simulate the D-layer, so the parameters are altered based on the F-layer data. The details will be elaborated in section \ref{sec53}. Moreover, the model is simulated with an actual beam whose gain pattern is different at each frequency. We investigate the change in the detected global 21-cm signal and foregrounds when subjected to the ionosphere. Also, chromaticity correction factors are applied when analysing the data.

\begin{figure}
    \centering
        \minipage{0.49\textwidth}
        \includegraphics[trim={0 0 3cm 0}, clip, width=\linewidth]{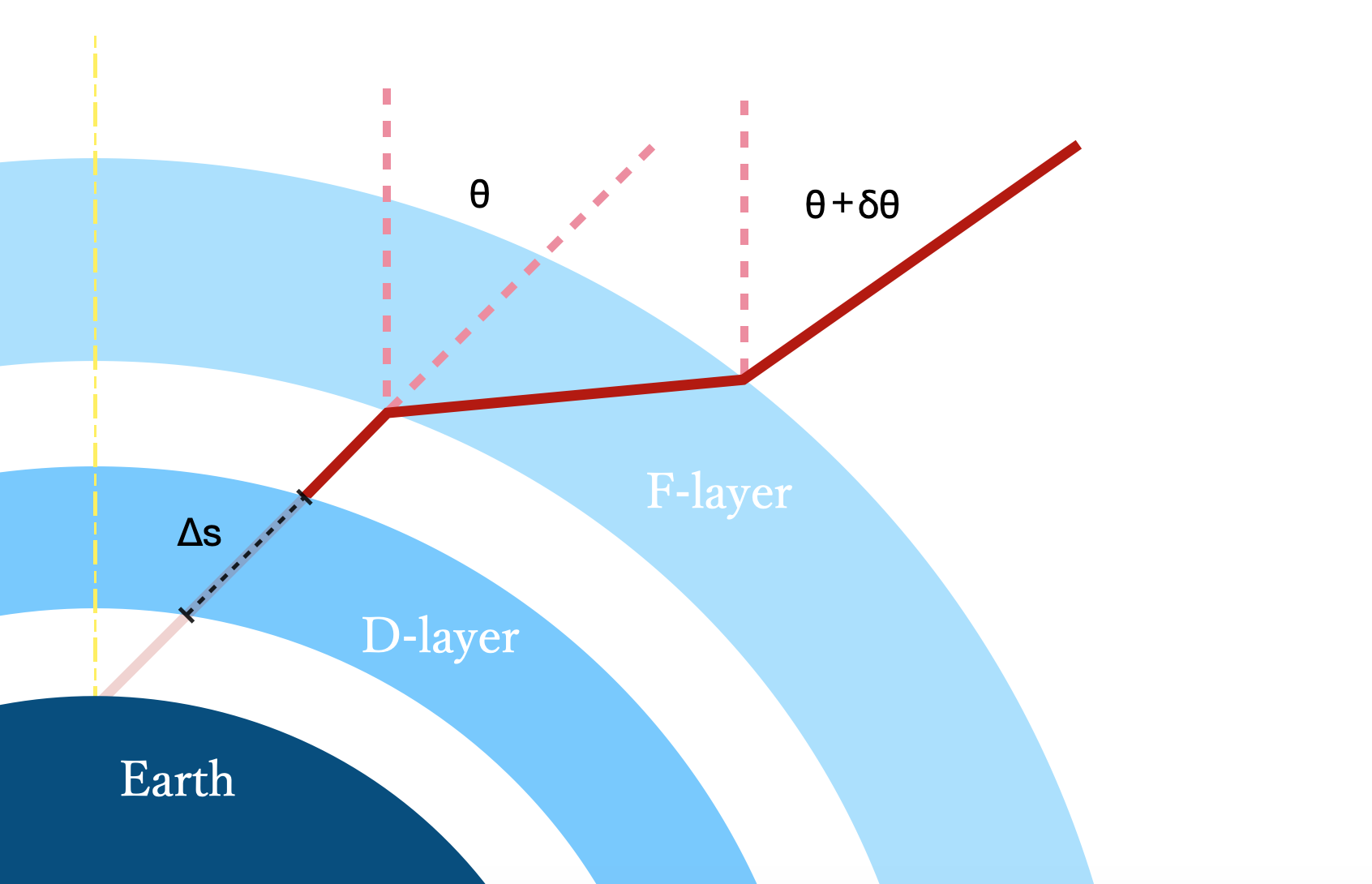}
        \endminipage\hfill
    \caption{Not-to-scale depiction of ionospheric refraction and absorption of homogeneous ionospheric layers \citep{km}.}
    \label{figioioio}
\end{figure} 
\subsection{Refractive Index}
The refractive index $\eta$ of the ionosphere is given by the generalised Appleton-Hartree equation \citep{shka}, in which $\eta$ is described by  electron density, magnetic field and wave propagation path. Because the change in refraction index due to the Earth’s magnetic field is insignificant \citep{km}, the magnetic field term is left out to simplify the complex calculation. The simplified Appleton-Hartree equation without the magnetic field term is:
\begin{equation}
\begin{aligned}
\eta^2 = 1 - \frac{(\nu_\mathrm{p}/\nu)^2}{1-i(\nu_\mathrm{c}/\nu)},
\end{aligned}
\end{equation}
where $\nu_\mathrm{p}$ is the electron plasma frequency, and $\nu_\mathrm{c}$ is the electron collision frequency. The electron plasma frequency can be determined by electron density $N_\mathrm{e}$ by the equation
\begin{equation}
\begin{aligned}
\nu_\mathrm{p} = \frac{1}{2\pi}\sqrt{\frac{N_\mathrm{e} e^2}{m_\mathrm{e} \epsilon_0}},
\end{aligned}
\end{equation}
where $e$ is the electron charge, $m_\mathrm{e}$ is the effective mass of the electron, and $\epsilon_0$ is the permittivity of free space. The electric field of a plane wave traveling in a homogeneous ionospheric layer is given by
\begin{equation}
\begin{aligned}
E(\Delta s) = E_0  \exp \left(-\frac{i2\pi \nu \Delta s}{c}\eta \right),
\end{aligned}
\label{efield}
\end{equation}
where $c$ is the speed of light, $\Delta s$ is the distance measured along the propagation path, and $E_0$ is the initial electric field. The real part of the refractive index depends mostly on the electron density; it introduces a change in the phase velocity, resulting in refraction. On the other hand, the imaginary part depends mostly on the electron collision rate, which is determined by the electron density, gas density and temperature. Being imaginary, it serves to dampen the wave amplitude exponentially, resulting in absorption.

The ionospheric model in this work is composed of two distinct parts (see Fig. \ref{figioioio}): one is the F-layer, accounting for refraction, and the other is the D-layer, accounting for absorption. The F-layer is characterised by low atmospheric gas density and high electron density, while the D-layer by high atmospheric gas density and low electron density. Thus, the F-layer is modelled with the real and frequency-dependent $\eta_F$, 
\begin{equation}
\begin{aligned}
\eta_F = \left( 1- \left( \nu_\mathrm{p} / \nu \right)^2 \right)^{1/2}, \;\; \nu_\mathrm{c} = 0,
\end{aligned}
\end{equation}
and the D-layer is modelled with the imaginary and also frequency-dependent $i\eta_D$,
\begin{equation}
\begin{aligned}
\eta_D \approx - \frac{1}{2} \frac{\nu_\mathrm{p}^2 \nu_\mathrm{c} / \nu}{\nu^2 + \nu_\mathrm{c}^2}.
\end{aligned}
\end{equation}
The parameters used to model the D-layer are drawn from a typical ionospheric condition \citep{theight,evans1968}.

\subsection{F-layer Refraction}
Two different equations used to describe electron density distribution within the F-layer have been tested in the model: the first one assumes a homogeneous F-layer shell spanning between 200 km and 400 km from the Earth's surface. The other one assumes an inhomogeneous layer at the same height, in which the electron density $N_\mathrm{e}$ varies parabolically with height, its peak lying in the middle of the layer \citep{bly}. The parabolic electron density as a function of height $h$ is then
\begin{equation}
\begin{aligned}
N_\mathrm{e}(h) = N_{0} \left[ 1 - \left( \frac{h-h_\mathrm{m}}{d}\right)^2 \right],
\end{aligned}
\label{elecdens}
\end{equation}
where $N_{0}$ is the maximum electron density at mean height $h_\mathrm{m}$, and $d$ is the semi-thickness of the layer. It, if modelled with the same maximum electron density, would, imaginably, result in a lesser degree of refraction than the homogeneous model. 

\begin{figure}
    \centering
    \minipage{0.49\textwidth}
        \includegraphics[width=\linewidth]{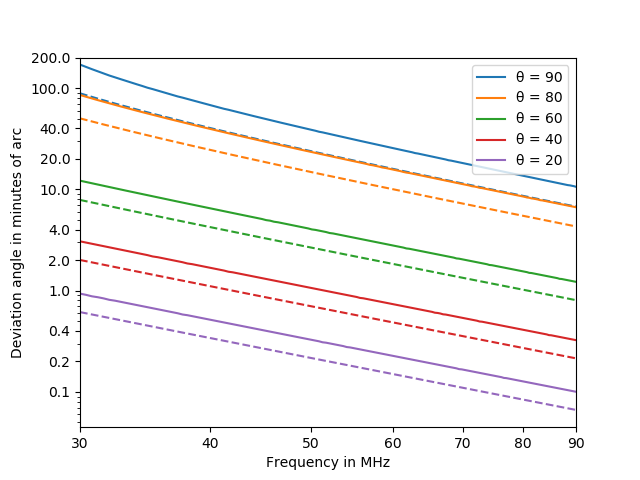}
    \endminipage\hfill
    \caption{Deviation angle of incoming rays at different zenith angles as a function of frequency. The dashed lines are simulated with a parabolic F-layer (equation \ref{elecdens}) and the solid lines are simulated with a homogeneous F-layer, recreated by using the same parameters suggested in \citet{km}: a homogeneous shell with (or the maximum electron density for the parabolic F-layer) $N_\mathrm{e} = 5.0\times 10^{11}\: \text{m}^{-3}$, and height 200-400 km from the Earth's surface. Propagation path increases with zenith angle. Therefore, incoming rays of a larger zenith angle suffer greater deviations.}
    \label{figdev}
\end{figure}

Due to the curvature of the Earth, the angle of the incoming ray is subject to deviation $\delta \theta$ while passing through the boundaries of the F-layer, whose total deviation can be calculated by Snell's law. For the parabolic layer, the total deviation experienced by an incoming ray of frequency $\nu$ can be written as
\begin{equation}
\begin{aligned}
\delta & \theta (\nu, \theta)   = \left(\frac{\nu_\mathrm{p}}{\nu}\right)^2 \frac{1}{d^2} a \sin{\theta}\int ^{h_\mathrm{m}-d}_{h_\mathrm{m}+d} \frac{(h-h_\mathrm{m})dh}{\eta_F^2[(h+a)^2\eta_F^2 - a^2\sin^2{\theta}]^\frac{1}{2}},
\end{aligned}
\label{dev}
\end{equation}
where $a$ is the Earth's radius. For the homogeneous model, this integral can be simplified as a discretised sum, considering only the contribution from the two surfaces on which $N_\mathrm{e}$ suffers a change. By Snell's law, any form of electron distribution can be easily calculated. Fig. \ref{figdev} shows the deviation angle of incoming rays of different incidence angles as a function of frequency, calculated by equation (\ref{dev}), as is done in \citet{km}.  

To not further complicate the matter at hand, for the rest of this paper, the simpler homogeneous F-layer is adopted. The parameters used to model the F-layer are  changed with time (every fifteen minutes) according to the data retrieved from Lowell GIRO Data Center.

The deviation angle will result in the change of direction for the incoming rays: it is shown in  \citet{km} that a chromatic increase in sky area of a few per cent, a disruption $\sim$ 2–3 orders of magnitude higher than the global signal, is introduced by the F-layer refraction. To model this effect, a \textit{lensing} effect is applied to the modified beam, and in so doing, the beam integrates to larger than unity over the entire sky. It also means different regions of the sky will be weighted differently according to their respective deviation angle. Some radiation close to the horizon coming from below may be able to acquire a new zenith angle smaller than $90^{\circ}$ after refraction, and hence could be captured by the antenna. The antenna temperature is expected to increase by $\sim$ 1-10 K due to refraction. The effective antenna beam 
\begin{equation}
\begin{aligned}
\hat{B}(\nu, \theta, \phi) = B(\nu, \theta + \delta\theta, \phi),
\end{aligned}
\end{equation}
due to the F-layer refraction can thus be written.

\section{D-layer Absorption}
The model assumes a homogeneous D-layer shell spanning between 60 km and 90 km from the Earth's surface with a constant electron density ${N}_{e} = 2.5 \times 10^{8}$ m$^{-3}$. The high atmospheric gas densities at these heights yield large electron collision frequencies that can result in significant absorption. The electron collision frequency is assumed to be 10 MHz \citep{nicole}. 
By equation (\ref{efield}), the amplitude of the electric field of a plane wave traveling in the D-layer is
\begin{equation}
\begin{aligned}
\lvert E(\Delta s)\rvert = E_0 \exp \left( \frac{2\pi \nu \Delta s}{c}\eta_D \right). 
\end{aligned}
\end{equation}
Since the intensity of an electromagnetic wave is proportional to the square of its amplitude, the loss factor that describes the loss of signal intensity can be written as
\begin{equation}
\begin{aligned}
 \mathcal{L}(\nu,\theta) = \exp \left( \frac{4\pi \nu \Delta s}{c}\eta_D \right).
\end{aligned}
\label{loss}
\end{equation}
The effective beam subject to absorption in the D-layer is then
\begin{equation}
\begin{aligned}
\hat{B}(\nu, \theta, \phi) = B(\nu, \theta, \phi)\mathcal{L}(\nu,\theta).
\end{aligned}
\end{equation}

\begin{figure}
    \centering
    \minipage{0.49\textwidth}
        \includegraphics[width=\linewidth]{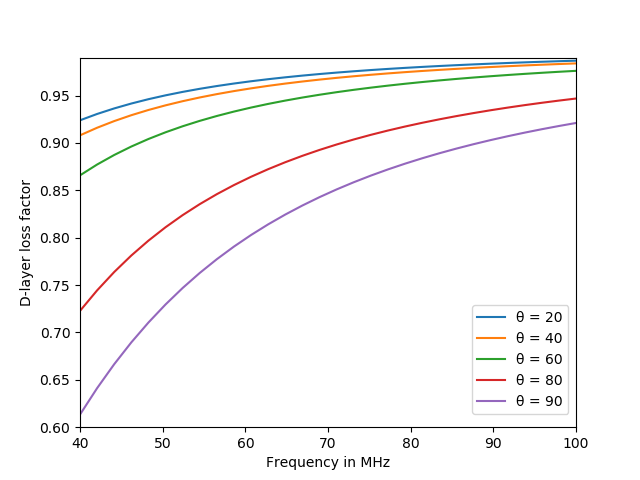}
    \endminipage\hfill
    \caption{Loss factor of incoming rays at different zenith angles as a function of frequency. The loss factor is calculated using the parameters: a homogeneous shell with $N_\mathrm{e} = 2.5\times 10^8\: \text{m}^{-3}$, $\nu_\mathrm{c} = 10$ MHz, and height 60-90 km from the Earth's surface.}
    \label{figloss}
\end{figure} 
Fig. \ref{figloss} shows the loss factor of incoming rays of different incidence angles as a function of frequency calculated by equation (\ref{loss}). While the total distance propagation path can be integrated numerically by Snell's law, it can also be approximated as
\begin{equation}
\begin{aligned}
\Delta s \approx \Delta h_D \left(1 + \frac{h_D}{a} \right)  \left(\cos^2{\theta} + \frac{2h_D}{a}  \right)^{-\frac{1}{2}},
\end{aligned}
\end{equation}
where $\Delta h_D$ is the width of the D-layer, and $h_D$ is the mean D-layer height. The absorption level in the relevant frequency range 50-200 MHz suggested by our model can be as large as $15\%$.

Combining the effects of both the F and the D-layer, the effective beam can be written:
\begin{equation}
\begin{aligned}
\hat{B}(\nu, \theta, \phi) = B(\nu, \theta + \delta\theta, \phi)\mathcal{L}(\nu,\theta).
\end{aligned}
\end{equation}
The simulated antenna temperature $T_\mathrm{A}$ can then be calculated by the modified integral,
\begin{equation}
\begin{aligned}
T_\mathrm{A}(\nu) = \int ^{2\pi}_{0}d\phi\int ^{\pi/2}_{0}d\theta \sin{\theta}T_\mathrm{f}(\nu,\theta,\phi)\hat{B}(\nu,\theta,\phi),
\end{aligned}
\label{anttemp}
\end{equation}
where $T_\mathrm{f}$ is the sky temperature at a given time.

\section{Computational and Analytical Methods}

\subsection{All-Sky Map}
In this paper, an all-sky map of base frequency $\nu_0 = 230$ MHz structured in \textsc{HEALpix} is used. The sky map is an instance of the 2008 Global Sky Model (\url{https://arxiv.org/abs/0802.1525}). It is scaled to a different frequency by the equation
\begin{equation}
\begin{aligned}
T_{\text{f},\nu} = \left(T_{\text{f},\nu_0}-T_{\text{CMB}}\right)\left(\frac{\nu}{\nu_0}\right)^{-\alpha} + T_{\text{CMB}},
\end{aligned}
\label{eqsky}
\end{equation}
where $T_{\text{CMB}} \sim 2.725 $ K is the average temperature of the cosmic microwave background radiation, and $\alpha$ is the spectral index. The spectral index varies across a realistic sky. However, the chromaticity function (equation \ref{ch1}) can correct the beam's chromaticity perfectly only when a constant spectral index is assumed. Therefore, in order to completely remove the chromatic effects introduced by the beam in the case where there is no ionosphere, with which the ionospherically compromised cases are compared, most of the cases shown in this paper will assume a constant spectral index $\alpha = 2.5$, a value that lies within a theoretical range derived by \citet{2017mnras}. Nevertheless, it is also critical to examine the signal in a more realistic scenario, so cases with a spatially varied set of spectral indices are also tested.

\subsection{Beam Chromaticity Correction}
The beam adopted in this model is the chromatic log-spiral antenna beam. In order to remove the chromaticity distortion from sky observations introduced by the chromatic antenna beam ${B}(\nu,\theta,\phi)$, the integrated antenna temperature (equation \ref{anttemp}) should be divided by a correction factor. The following equation for such a factor is suggested in \citet{2017mnras}:
\begin{equation}
\begin{aligned}
C(\nu) = \frac{\int ^{2\pi}_{0}d\phi\int ^{\pi/2}_{0}d\theta T_\mathrm{f}(\nu_0,\theta,\phi){B}(\nu,\theta,\phi)}
{\int ^{2\pi}_{0}d\phi\int ^{\pi/2}_{0}d\theta T_\mathrm{f}(\nu_0,\theta,\phi){B}(\nu_0,\theta,\phi)},
\end{aligned}
\label{ch1}
\end{equation}
where $\nu_0$ denotes the reference frequency. This correction factor works under the assumption that the spectral index is constant across the entire sky. Moreover, the constant $T_{\text{CMB}}$ in equation (\ref{eqsky}), if not removed from the sky temperature, would introduce a component that does not scale to the frequency dependent power law. Therefore, in order to remove the beam chromaticity perfectly in our model, the following equation is adopted:
\begin{equation}
\begin{aligned}
C(\nu) = \frac{\int ^{2\pi}_{0}d\phi\int ^{\pi/2}_{0}d\theta [T_\mathrm{f}(\nu_0,\theta,\phi)-T_{\text{CMB}}]{B}(\nu,\theta,\phi)}
{\int ^{2\pi}_{0}d\phi\int ^{\pi/2}_{0}d\theta [T_\mathrm{f}(\nu_0,\theta,\phi)-T_{\text{CMB}}]{B}(\nu_0,\theta,\phi)}.
\end{aligned}
\label{ch12}
\end{equation}
 The difference between the integrated antenna temperature corrected by equation (\ref{ch1}) and (\ref{ch12}) exceeds 0.3 K at lower frequencies, which is significant enough to corrupt a signal as weak as the global 21-cm line. This difference is also apparently chromatic.

A similar modification should apply also to the antenna temperature integration; it should be subtracted before it is corrected by the chromaticity correction, and then added back afterwards:
\begin{equation}
\begin{aligned}
&T_\mathrm{A}(\nu) 
\\&=  T_{\text{CMB}} + \frac{1}{C(\nu)} \int ^{2\pi}_{0}d\phi\int ^{\frac{\pi}{2}}_{0}d\theta \sin{\theta}[T_\mathrm{f}(\nu,\theta,\phi)-T_{\text{CMB}}]{B}(\nu,\theta,\phi).
\end{aligned}
\label{anttemp2}
\end{equation}

When these assumptions hold, the integrated antenna temperature at each frequency will be readjusted towards the corresponding beam gain at the chosen reference frequency, and it will remove the chromatic effect introduced by the originally chromatic beam. In order to see how well this equation could apply to the chromatic ionospheric effects, the modified equation is also tested:
\begin{equation}
\begin{aligned}
C_{\mathrm{ion}}(\nu) = \frac{\int ^{2\pi}_{0}d\phi\int ^{\pi/2}_{0}d\theta T_\mathrm{f}(\nu_0,\theta,\phi)\hat{B}(\nu,\theta,\phi)}
{\int ^{2\pi}_{0}d\phi\int ^{\pi/2}_{0}d\theta T_\mathrm{f}(\nu_0,\theta,\phi)\hat{B}(\nu_0,\theta,\phi)},
\end{aligned}
\label{ch2}
\end{equation}
where an ionospherically affected beam is adopted:
\begin{equation}
\begin{aligned}
\hat{B}(\nu, \theta, \phi) = B(\nu, \theta + \delta\theta, \phi)\mathcal{L}(\nu,\theta).
\end{aligned}
\label{eqbeamm}
\end{equation}
In the cases where simulations span over a period of time, the chromaticity function is calculated and applied at each time step before time integration.

It should be noted, however, that this simple correction is only used in this paper to demonstrate the degree of change the ionosphere can cause. In practice, the REACH experiment \citep{reachh} adopts a model using full Bayesian pipeline technique that allows a more sophisticated parameter estimation \citep{anstey}. 

\subsection{Data Structure and Gridding}
The total antenna temperature is integrated in the ring pixel scheme adopted in the \textsc{HEALpix} coordinate instead of the conventional spherical coordinates suggested in equation (\ref{anttemp}). In our model, there are $12 \times 512^2$ (${N}_{\text{side}}$ $= 512$) pixels in total, corresponding to a resolution of $\sim 0.002$ in radian, or $\sim 6.87$ minutes of arc. High resolution is required for this model, as the F-layer deviation can be very small at small zenith angle and at higher frequencies, as shown in Fig. \ref{figdev}. 

While the sky map is stored in the \textsc{HEALpix} structure, the log-spiral beam is designed in spherical polar coordinates, and thus needs to be transformed and remapped into the \textsc{HEALpix} structure by interpolation before the pixel-wise multiplication of the sky map and the effective antenna beam (equation \ref{eqbeamm}). Before being subjected to the ionospheric expansion imposed by the F-layer, the beam is renormalised to make sure the beam integrates to unity over the entire sky at every frequency.
 
\subsection{Modelled Global 21-cm Signal}
In the simulations presented in this work, a 21-cm signal, assumed to be a normal Gaussian: 
\begin{equation}
T_{21}(\nu) = A_0\:e^{\frac{-(\nu-\mu)^2}{2\sigma^2}},
\end{equation}
is inserted to the sky map to test the effect the ionospheric disturbances have on the existing method to extract the signal. The modelled 21-cm signal is assumed to be constant across the sky with no spatial variation. Gaussian signals motivated by  \citet{edges} as well as those of various amplitudes $A_0$, central frequencies $\mu$, and width $\sigma$ are tested. The sky temperature with an inserted global 21-cm signal is given by
\begin{equation}
T_{\text{sky}}(\nu,\theta,\phi) = T_\mathrm{f}(\nu,\theta,\phi) + T_{21}(\nu),
\end{equation}
where $T_\mathrm{f}(\nu,\theta,\phi)$ is $T_{\mathrm{f},\nu}$ in equation (\ref{eqsky}). This is the sky temperature we use in the model when simulating with a global 21-cm signal. 

\subsection{Data Analysis}
\label{secda}
It is essential to know the change in the detected antenna temperature the ionospheric effects would introduce, for it is what affects the values we get from direct observations.

As the simulation considers a ionosphere that varies in time, it should be noted that the change in antenna temperature  is dependent on the sky region at which the beam is pointing, as well as the ionospheric condition at the time of observation. It is speculated that the chromatic ionospheric disturbances would cancel out over time, so, in such cases, in order to make a fair comparison between different integration periods, the integrated antenna temperature needs to be scaled to a common ground. 

A similar problem arises when it comes to the chromaticity correction mentioned earlier. Equation (\ref{ch1}) shows that when this chromaticity correction is applied, it is corrected with respect to the beam gain $B(\nu_0, \Omega)$ at the reference frequency $\nu_0$, where $\Omega$ is the angular position of the sky area at which the beam is pointing, which means the final product of this correction will differ according to the reference frequency; if the beam gain $B(\nu, \Omega)$ at $\nu_1$ is smaller than at $\nu_2$ by a factor of $a$ at $\Omega$, $a(\Omega) B(\nu_1, \Omega) =B(\nu_2, \Omega)$, with $\int B(\nu, \Omega)d\Omega = 1$ for all frequencies:
\begin{equation}
\begin{aligned}
C_{\nu_2}(\nu) \approx \frac{1}{A} \times \frac{\int T_\mathrm{f}(\nu_1,\Omega)  {B}(\nu,\Omega) d\Omega}{\int T_\mathrm{f}(\nu_1,\Omega) {B}(\nu_1,\Omega)d\Omega}  = \frac{C_{\nu_1}(\nu)}{A},
\end{aligned}
\label{sca}
\end{equation}
where $A$ is a constant.

As shown in the derivation, the difference is not a simple constant but a factor convolved with $T_{\text{CMB}}$ and it is also dependant on the angular location. This correction is adopted in some of the calculations just to show a fairer comparison when a different reference frequency is applied, but a fair comparison might not be needed in practice since the raw, unscaled, value is perhaps what really matters in the experiment.

Besides the difference in magnitude, it is perhaps more critical to see how the chromatic profile is altered; it is the intermixed chromatic effect that confuses the signal measurement. As the foreground can be said to obey a power law $T_\text{sync} \propto \nu^{-\alpha}$, the $n^{\text{th}}$ order log-polynomial fitting of the following form is adopted to fit the foregrounds:
\begin{equation}
\begin{aligned}
T_\mathrm{M}(\nu) = T_{\text{CMB}} + 10^{\sum^n_{i=0} a_i \log(\nu)^i }.
\end{aligned}
\label{logpoly}
\end{equation}
In this paper, we adopt a $5^{\text{th}}$ log-polynomial to fit the data, as including higher order terms would risk modelling out the cosmological signal. This technique has been proposed in the literature \citep{Pritch2008,hark2012} and is widely used in analysing global signal. However, the merit of this kind of fitting works well under the assumption of a smooth foreground; it falls short when a chromatic beam is used, where the spatial variation in the sky temperature is coupled into the spectral structure, especially for a wide frequency band. necessarily imply a 

Since the simulations include a signal in the form of a Gaussian, the complete form of the model that is used to fit the simulated data is:
\begin{equation}
\begin{aligned}
T_\mathrm{M}(\nu) = T_{\text{CMB}} + 10^{\sum^n_{i=0} a_i \log(\nu)^i }+ A_0\:e^{\frac{-(\nu-\mu)^2}{2\sigma^2}} ,
\end{aligned}
\label{logpolygau}
\end{equation}
with constraints on each parameter, the values of which depend on the signal added in the simulation. If our model for the foregrounds is accurate enough, the latter form (equation \ref{logpolygau}) should fit the simulated data better than the previous form (equation \ref{logpoly}) that only takes the foregrounds into account. Both forms of fitting will be shown in section \ref{sec5}. Also, one must stress that the yielded residuals after fitting do not necessarily imply an ionospheric feature, as any distortion to a originally smooth foreground would introduce an undulation to the residual of a log-polynomial fit.

\section{Simulation Results}
\label{sec5}
In this section, the plots and analyses of the simulated data generated by the aforementioned model will be shown. The effects coming from the two different layers will be looked into separately before moving on to the simulations where both layers are in effect.

\subsection{Separate Effects}
\begin{figure}
    \centering
    \minipage{0.532\textwidth}
        \includegraphics[trim={0.21cm 0.2cm 0 0},clip,width=\linewidth]{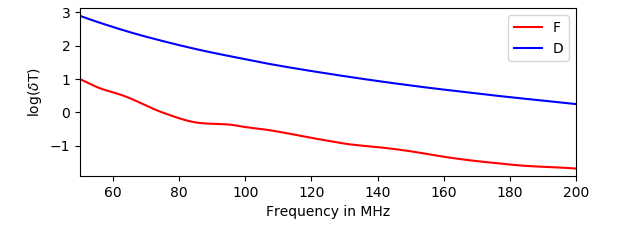}
    \endminipage\hfill
    \caption{Difference in integrated antenna temperature when only the F-layer (red) or the D-layer (blue) is considered, corrected by equation (\ref{ch1}) with reference frequency $\nu_0 = 150$ MHz. It is shown in the absolute difference (with respect to the case without an ionosphere) on a log scale. The F-layer increases while the D-layer decreases the overall antenna temperature. The D-layer (absorption) dominates the ionospheric effect in magnitude, about 15 times that of the F-layer (refraction).}
    \label{figfd}
\end{figure}

Fig. \ref{figfd} shows that the F-layer alone increases the foreground temperature by $\sim 10$ K at 50 MHz, resulting in a $\sim 0.2\%$ increment compared to the original antenna temperature, while the D-layer reduces the signal by $\sim 800$ K at 50 MHz, resulting in a $\sim 16\%$ deficit compared to the original antenna temperature. The D-layer dominates the ionospheric effect in magnitude, about 15 times that of the F-layer. However, we are more interested in how it affects the \textit{shape}, which, when unaffected by the ionosphere or other effects, should resemble the power law that describes the foregrounds dominated by synchrotron radiation.

\begin{figure*}
    \centering
    \minipage{0.5\textwidth}
        \includegraphics[trim={0 0.7cm 0 6.48cm},clip,width=\linewidth]{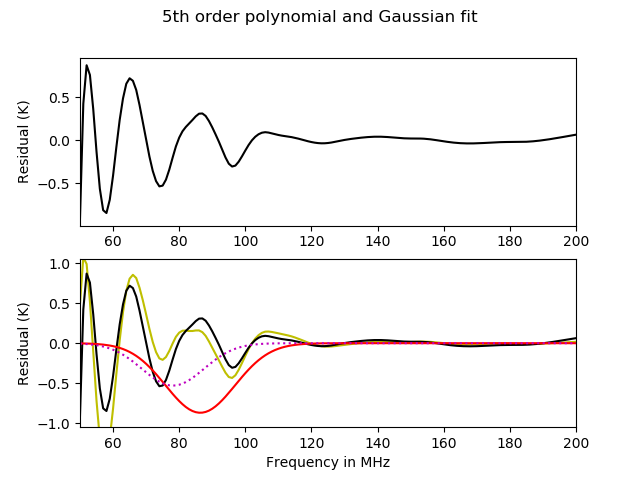}
    \endminipage\hfill
    \minipage{0.5\textwidth}
        \includegraphics[trim={0 0.7cm 0 6.48cm},clip,width=\linewidth]{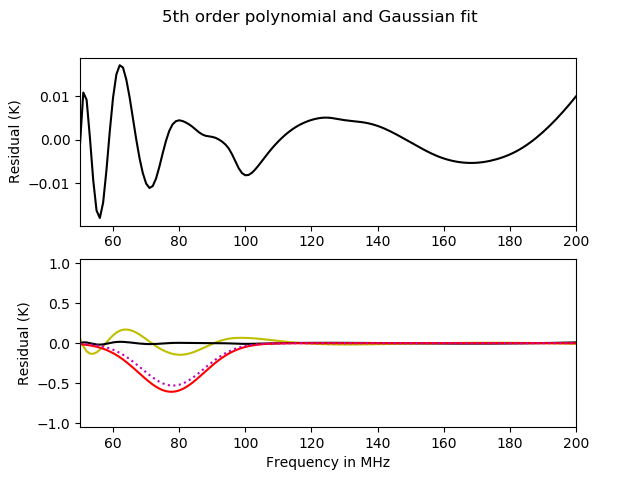}
    \endminipage\hfill

    \centering
    \minipage{0.5\textwidth}
        \includegraphics[trim={0 0 0 6.48cm},clip,width=\linewidth]{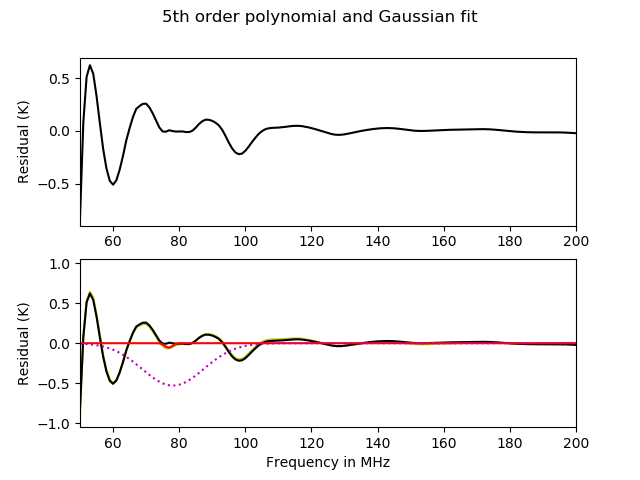}
    \endminipage\hfill
    \minipage{0.5\textwidth}
        \includegraphics[trim={0 0 0 6.48cm},clip,width=\linewidth]{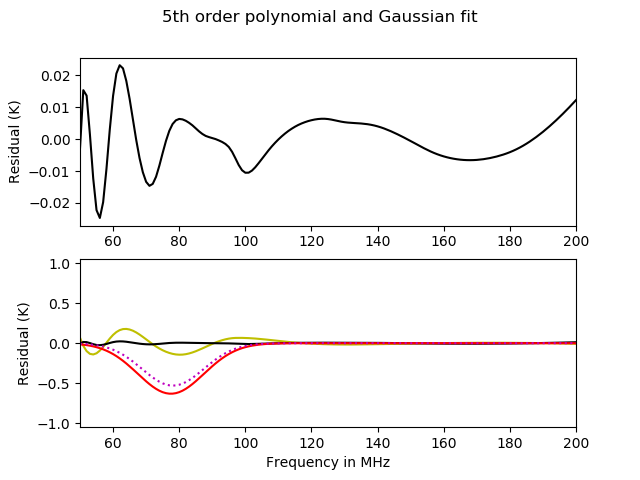}
    \endminipage\hfill
    \caption{The plots show the residuals of the signal modelled with ionospheric effects, fitted by equation (\ref{logpoly}) and (\ref{logpolygau}) using least-squares fit. Simulated with \textbf{only} the (top-left) D-layer, a snapshot,  (top-right) F-layer, a snapshot, (bottom-left)  D-layer, constant in time, integrated every 15 minutes over 5 nights,  (bottom-right) F-layer, time-varying, integrated every 15 minutes over 5 nights. We show: \textbf{black} -- residual of the least-squares fit of a $5^{\text{th}}$ order log-polynomial and a Gaussian (equation \ref{logpolygau}), \textbf{olive} -- residual of the least-squares fit of a $5^{\text{th}}$ order log-polynomial only (equation \ref{logpoly}), \textbf{red} -- the Gaussian component of the combined least-squares fit (equation \ref{logpolygau}), and \textbf{dotted-magenta} -- the 21-cm signal added in the simulation.  Greater residuals and more ill-fitted signals are yielded in the D-layer cases, suggesting that the D-layer has an greater overall effect on the chromatic variance.}
    \label{figrfd}
\end{figure*}

Fig. \ref{figrfd} shows the residual after being fitted using equation (\ref{logpoly}) and (\ref{logpolygau}). The fitted Gaussian signal is closer to the simulated signal in terms of amplitude and the central frequency for the case where only the F-layer is simulated. Moreover, the residual of the F-layer is weaker than that of the D-layer. It  suggests that the D-layer has a greater effect on the overall chromatic variance of the integrated antenna temperature.
In the presence of only the F-layer, a global signal of amplitude $\sim 0.5$ K is still detectable by the simple  log-polynomial fitting (equation \ref{logpolygau}). However, in other cases simulated with a different set of parameters, where the residual yielded by the fitting exceeds 0.2 K, the F-layer alone can introduce significant distortions to the foregrounds. Although the signal in this simulation has bigger amplitude, $\sim 0.5$ K, in reality, it is possible that the signal is as small as 0.1 K \citep{cohen2017}. In such cases, a $0.2$ K disturbance can still overwhelm the signal. 

Fig. \ref{figrfd} also shows the residual of the data integrated over a longer time interval (five nights), and a longer integration time appears to reduce the residual in the D-layer case. However, it is most likely due to the lesser effect at other time-steps that the overall distortion averages out when integrated, and this result is therefore not sufficient enough to be taken as an evidence that the ionospheric effect cancels out over time. The discussion will be elaborated in section \ref{sec53}.

\begin{table}
	\centering
	\caption{The table shows the root mean square (RMS) of the residuals after fitting and the amplitude $|A_0|$, central frequency $\mu$, and width  $\sigma$ (standard deviation) of the fitted Gaussian signal. The reference (in red) is the original input signal used in the model. The larger residuals in the D-layer cases suggest that it contributes more to the ionospheric distortion on the detected signal.}
	\label{tabrfd}
	\begin{tabular}{lcccc} 
		
		&Residual& \multicolumn{3}{c}{Fitted Gaussian Signal} \\
		\hline
		 & RMS (K) & $|A_0|$ (K) & $\mu$ (MHz) & $\sigma$ (MHz)\\
		\hline
		\textcolor{BrickRed}{Reference} & -- & \textcolor{BrickRed}{0.53} &\textcolor{BrickRed}{78.1} &\textcolor{BrickRed}{18.7}\\
		D-layer, snapshot & $4.08 \times 10^{-1}$ & 0.87 & 86.37 & 21.0\\
		D-layer, 5 nights & $2.49 \times 10^{-1}$ & 0.05 & 76.9 & 2.23\\
		F-layer, snapshot & $7.80 \times 10^{-3}$ & 0.61 & 77.6 & 19.8\\
		F-layer, 5 nights & $1.05 \times 10^{-2}$ & 0.63 & 77.5 & 20.0\\
		\hline 
	\end{tabular}
\end{table}

\subsection{Combined Effect}
 To get a grasp of the range of the total ionospheric effect can have on the integrated antenna temperature within the ionospheric parameter space, the parameters including electron density $N_\mathrm{e}$, electron collision frequency $\nu_\mathrm{c}$, height, and thickness, twelve sets of randomised parameters within respective realistic ranges in a normal ionospheric weather were simulated. This result suggests that the integrated antenna temperature subjected to ionospheric distortions modelled within a given scope can have a 2500 K difference at 50 MHz and the range narrows as the frequency increases. The percentage of difference also decreases with growing frequency.

\begin{figure}
    \minipage{0.532\textwidth}
        \includegraphics[trim={0.21cm 0 0 6.48cm},clip,width=\linewidth]{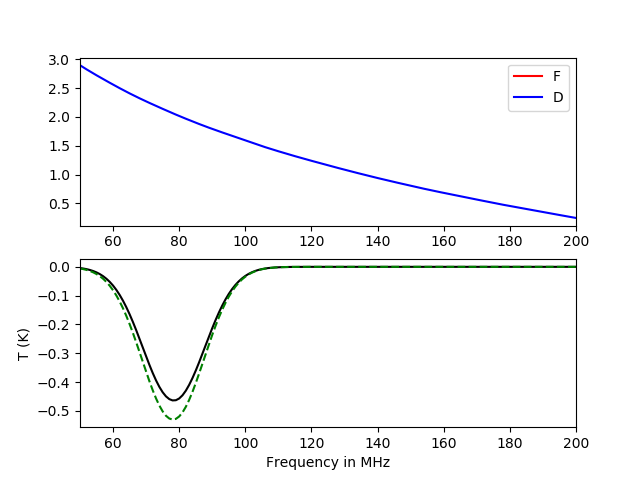}
    \endminipage\hfill
    \caption{Change in the global 21-signal due to the ionosphere. The parameters used to model the ionosphere are randomised within the theoretical ranges for a normal ionospheric weather. The plot shows the inserted signal in  the dotted green line and the ionospherically affected signal in the solid black line. In this case, the amplitude is weakened by $\sim 15\%$. One can also observe a chromatic distortion in the signal, that it suffers a greater change at lower frequencies.}
    \label{fig21change}
\end{figure}

It is also critical to know how the inserted global 21-cm signal could be affected due to the ionosphere. Fig. \ref{fig21change} shows the change due to the ionosphere simulated using randomised parameters. The plots show that, for a signal centring at 78.1 MHz, the amplitude can be weakened by $\sim 15\%$, which is similar to the D-layer absorption rate shown in Fig. \ref{figfd}. The change in the Gaussian shape  shows the chromatic distortions introduced by the ionosphere, whose effects increase with lowering frequency.

\subsection{Different Integration Time Length}
\label{sec53}
A case with a constant and another with a time-varying D-layer, both with a time-varying F-layer, are shown in this section. There is currently no available data required to simulate the D-layer. Nevertheless, we are still interested in investigating the behaviour of a time-varying D-layer. To achieve that, we instead change the D-layer parameters according to the actual time-varying F-layer data by scaling, so that each scale factor of the same parameter of the D-layer and the F-layer corresponds to the respective scale factor suggested in the theoretical values adopted in \citet{km} (the same values are also adopted in Fig. \ref{figdev} and \ref{figloss}). We have also made sure that those values lie within the theoretical ranges. Since electron collision frequency $\nu_c$ is considered null when modelling the F-layer, a constant $\nu_\mathrm{c} = 10$ MHz is adopted for the D-layer. This is only an attempt to study a varying D-layer, and may not be representative of an actual ionosphere. 

\begin{figure}
    \centering
        \minipage{0.532\textwidth}
        \includegraphics[trim={0.21cm 0.7cm 0 0},clip,width=\linewidth]{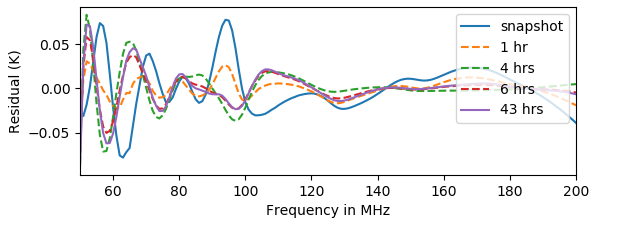}
        \endminipage\hfill
        \minipage{0.532\textwidth}
        \includegraphics[trim={0.21cm 0.2cm 0 0},clip,width=\linewidth]{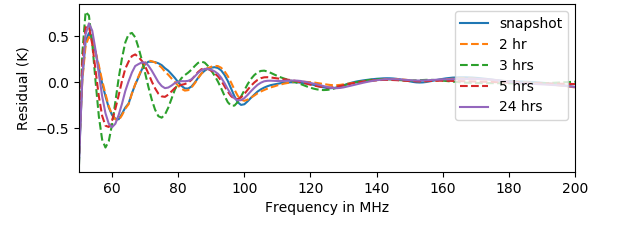}
        \endminipage\hfill
    \caption{Residuals of the least-squares fit of a $5^{\text{th}}$ order log-polynomial and a Gaussian (equation \ref{logpolygau}) to data integrated every 15 minutes over different lengths of time. The \textbf{upper} panel is simulated with a time-varying D-layer and F-layer, while the \textbf{lower} panel is simulated with a constant D-layer and a time-varying F-layer. Those integrated over a longer time do not yield a weaker residual or a significantly better fit, showing no clear sign that the chromatic ionospheric effects cancel over time in both cases.}
    \label{figintttt3}
\end{figure}

Fig. \ref{figintttt3} shows the residuals of the data integrated over different lengths of time. It would be convenient if the chromatic ionospheric effects cancel over a long integration time; in such a case, the ionosphere can be ignored simply by temporal integration. However, by the way the data is analysed and fitted in this work (equation \ref{logpolygau}), there is no clear sign showing the cancellation of chromatic ionospheric effects over time for both constant and time-varying D-layer; the residual is still comparable to the global 21-cm signal in magnitude even after integration over five nights. The absence of this property suggests that an alternative approach to remove this significant distortion to the foregrounds should be proposed to accurately measure the global 21-cm signal.

One can also observe that the residuals decrease by about an order of magnitude in the case of a time-varying D-layer. It is due to a milder ionospheric condition compared to the theoretical values used to simulate a constant D-layer. It suggests that the significance of the ionospheric effects depends greatly on the current ionospheric condition. 

\begin{table}
	\centering
	\caption{The table shows the root mean square (RMS) of the residuals after fitting and the amplitude $|A_0|$, central frequency $\mu$, and width  $\sigma$ (standard deviation) of the fitted Gaussian signal. The reference (in red) is the original input signal used in the model. The residual does not have a consistent decrease as the integration time period increases.}
	\label{tabpap}
	\begin{tabular}{rcccc} 
		
		&Residual& \multicolumn{3}{c}{Fitted Gaussian Signal} \\
		\hline
		 & RMS (K) & $|A_0|$ (K) & $\mu$ (MHz) & $\sigma$ (MHz)
		\\
		\hline 
		\textcolor{BrickRed}{Reference} & -- & \textcolor{BrickRed}{0.53} &\textcolor{BrickRed}{78.1} &\textcolor{BrickRed}{18.7}\\
		[-0.0ex] \multicolumn{5}{l}{{time-varying D-layer}}\\
		
		\hline
		snapshot & $3.92 \times 10^{-2}$ & 0.21 & 81.3 & 7.38\\
		1 hour & $1.41 \times 10^{-2}$ & 0.22 & 80.7 & 9.86\\
		4 hours & $3.40 \times 10^{-2}$ & 0.38 & 79.6 & 15.6\\
		6 hours & $2.45 \times 10^{-2}$ & 0.30 & 79.7 & 13.5\\
		43 hours & $3.01 \times 10^{-2}$ & 0.27 & 79.1 & 12.2\\
		\hline \\[-1.5ex] \multicolumn{5}{l}{{constant D-layer}}\\
		
		\hline
		snapshot & $3.51 \times 10^{-1}$ & 0.68 & 89.3 & 23.8\\
		2 hour & $2.25 \times 10^{-1}$ & 0.45 & 75.9 & 41.4\\
	    3 hours & $3.35 \times 10^{-1}$ & 0.71 & 88.3 & 25.0\\
		5 hours & $2.39 \times 10^{-1}$ & 0.37 & 90.0 & 25.0\\
		24 hours & $2.45 \times 10^{-1}$ & 0.46 & 82.0 & 40.0\\
		\hline 
	\end{tabular}
\end{table}

\subsection{Pattern of the Residuals}

\begin{figure}
    \minipage{0.532\textwidth}
        \includegraphics[trim={0.21cm 0.2cm 0 0.9cm},clip,width=\linewidth]{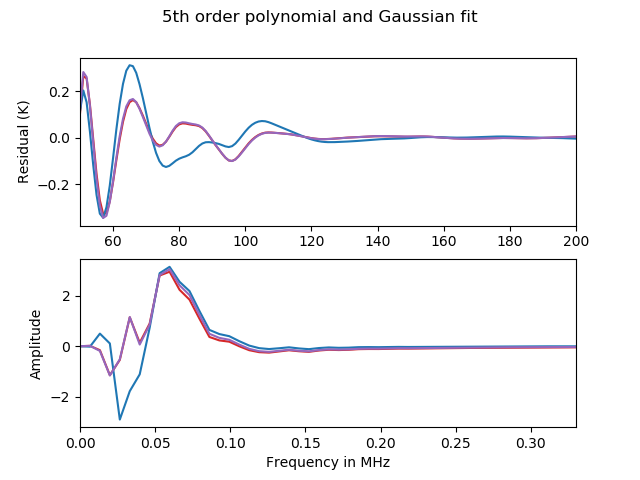}
    \endminipage\hfill \\
    \caption{Residuals (upper panel) and their Fourier components (lower panel), (purple) simulated with the D-layer, the F-layer, and a global 21-cm signal, (red) simulated with the D-layer, and the F-layer, (blue) simulated with the D-layer only. The frequencies at which the local maxima and minima of the residuals are situated are consistent in all cases. The amplitude fluctuation with frequency is also consistent. This pattern is dominated by the D-layer, with the F-layer having a minor effect.}
    \label{figrandrannran}
\end{figure}

When simulated without a global 21-cm signal, the pattern of the residual after log-polynomial fitting (equation \ref{logpoly}) is highly consistent in its shape in several different cases. This pattern is shown in Fig. \ref{figrandrannran} to be dominated by the D-layer, with the F-layer having a minor effect. Fig. \ref{figrandrannran} shows the residuals after log-polynomial fitting to the data simulated with six sets of randomised D-layer parameters and without the F-layer. The frequencies at which the local maxima and minima of the residuals are situated are consistent in all cases. The amplitude fluctuation with frequency is also consistent. The physics underlying this pattern and its mathematical composition is still unclear. 

This characteristic motivates us to modify the fitting equation (equation \ref{logpolygau}) by adding a term that describes this pattern, so it becomes:
\begin{equation}
\begin{aligned}
T_\mathrm{f}(\nu) = T_{\text{CMB}} + 10^{\sum^n_{i=0} a_i \log(\nu)^i }+ A_0\:e^{\frac{-(\nu-\mu)^2}{2\sigma^2}} + B_0 R_{D}(\nu),
\end{aligned}
\label{logpolygaufff}
\end{equation}
where $B_0$ is a constant and $R_D(\nu)$ is the pattern as a function of frequency. However, the results (see Table \ref{tabfff1}) are only improved when the original fits are good in the first place; in the cases where the signals are already discernible by using equation (\ref{logpolygau}), the residuals are significantly reduced by using the modified equation. This equation is applicable only to data for a single time step. When applying the same method to the temporally integrated cases, this method fails to generate a proper result. This suggests that different $B_0$'s in each time step should be calculated separately and cannot be simplified by one $B_0$ after the data are integrated over time. An alternative way to get round this issue is to subtract this term from the data at each time step before time integration. However, the Gaussian fitting and the foreground fitting should be done in one single process, and doing so, extracting the Gaussian signal at every time step, would therefore run counter the purpose of time integration, which is to let the chromatic ionospheric effect cancel itself out over time before signal extraction.

\begin{table}
	\centering
	\caption{The table shows the root mean square (RMS) of the residuals after fitting and the amplitude $|A_0|$, central frequency $\mu$, and width  $\sigma$ (standard deviation) of the fitted Gaussian signal. The reference (in red) is the original input signal used in the model. The same sets of data are fitted using two different equations, one is fitted with the additional D-layer pattern term. Some cases show improvements when the D-layer pattern is also being fitted while some yield a null fitted signal.}
	\label{tabfff1}
	\begin{tabular}{rcccc} 
		
		&Residual& \multicolumn{3}{c}{Fitted Gaussian Signal} \\
		\hline
		 & RMS (K) & $|A_0|$ (K) & $\mu$ (MHz) & $\sigma$ (MHz) \\
		\hline 
		\textcolor{BrickRed}{Reference} & -- & \textcolor{BrickRed}{0.53} &\textcolor{BrickRed}{78.1} &\textcolor{BrickRed}{18.7}
		\\[-0.0ex] \multicolumn{5}{l}{{Fitted without the D-layer Pattern  (equation \ref{logpolygau})}}\\
		
		\hline
		snapshot 1 & $2.35 \times 10^{-1}$ & 0.45 & 80.8 & 39.6\\
		snapshot 2 & $4.12 \times 10^{-1}$ & 0.99 & 88.6 & 25.6\\
		snapshot 3 & $3.25 \times 10^{-1}$ & 0.99 & 85.9 & 27.0\\
		2 hours & $2.25 \times 10^{-1}$ & 0.45 & 75.9 & 41.4\\
		24 hours & $2.45 \times 10^{-1}$ & 0.46 & 82.0 & 40.0\\

		\hline \\[-1.5ex] \multicolumn{5}{l}{{Fitted with the D-layer Pattern  (equation \ref{logpolygaufff})}}\\
		
		\hline
		snapshot 1 & $2.23 \times 10^{-1}$ & 0.00 & 79.7 & 41.6\\
		snapshot 2 & $2.10 \times 10^{-2}$ & 0.45 & 78.2 & 19.7\\
		snapshot 3 & $1.67 \times 10^{-2}$ & 0.20 & 82.0 & 11.5\\
		2 hours & $2.15 \times 10^{-1}$ & 0.00 & 64.0 & 38.6\\
		24 hours & $9.18 \times 10^{-2}$ & 0.99 & 61.9 & 7.22\\

		\hline 
	\end{tabular}
\end{table}

\subsection{Modified Beam Chromaticity Correction and Spatially Varying Spectral Index}
We are interested in knowing whether the modified chromaticity correction that bears a similar form to equation (\ref{ch12}), used to correct the beam chromaticity, can also help mitigate the chromaticity imposed by the ionosphere. Technically, it is correcting for both the ionosphere and the beam, for the two are convolved for cases simulated with an ionosphere. 

\begin{figure}
    \centering
    \minipage{0.532\textwidth}
        \includegraphics[trim={0 5.2cm 0 0.9cm},clip,width=\linewidth]{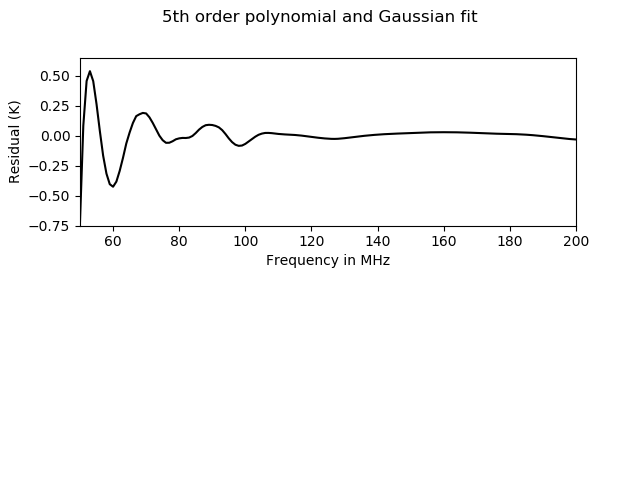}
    \endminipage\hfill
    \caption{Residual when the beam chromaticity is corrected by equation (\ref{ch2}) (integrated over 5 night, 24 hours). Data corrected by the modified equation give a slightly lower residual than by the original equation (\ref{ch12}).}
    \label{figsicsic}
\end{figure}

The difference between the data corrected by the two different equations is small, $\sim$ 0.1 K at lower frequencies. Although the magnitude of this difference is still comparable to the brightness of a global 21-cm signal, the difference does not decrease with frequency like a power law, but seems to fluctuate about 0 K. The residuals are similar  in both cases, with the one corrected by the modified equation having a slightly smaller residual (Fig. \ref{figsicsic}), suggesting that the modified equation has more or less improved the result. However, more cases need to be tested in order to ascertain the effectiveness of the modified correction equation.

We have also studied the residual of a case simulated with a uniform spectral index $\alpha = 2.5$ and another with realistic spectral indices that vary across the sky. The map with realistic spectral indices is built by interpolating the GSM and the Haslam map \citep{anstey}. As the adopted chromaticity correction (equation \ref{ch12}) only works under the assumption that the spectral index is uniform, it is expected that the correction would fail to effectively correct for the coupled chromatic disruption caused by the antenna beam itself and the ionospheric effects when the spectral index varies spatially.

\subsection{Different Reference Frequency}
We study the difference between two sets of data simulated with the same ionosphere, but whose beam/ionosphere chromaticity is corrected by equation (\ref{ch2}) using different reference frequencies. When doing a magnitude comparison, they are also corrected by a factor calculated by equation (\ref{sca}) to compensate for the different reference temperature at different frequency caused by a chromatic beam. 

\begin{figure}
    \centering
        \minipage{0.532\textwidth}
        \includegraphics[trim={0.21cm 0.2cm 0 0},clip,width=\linewidth]{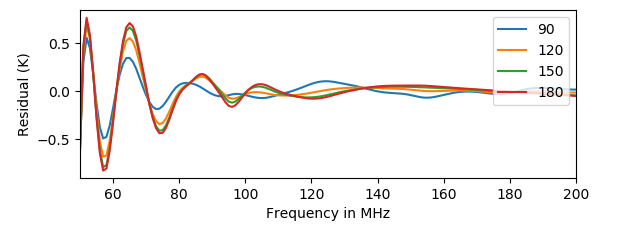}
        \endminipage\hfill
    \caption{Residuals of the least-squares fit of a $5^{\text{th}}$ order log-polynomial and a Gaussian (equation \ref{logpolygau}) to data corrected using different reference frequencies: 90 MHz, 120 MHz, 150 MHz, and 180 MHz. The difference in the residuals can be observed. The chromatic beam no longer integrates to unity when coupled with the chromatic ionospheric effects, and thus introduces different distortions at different frequencies.}
    \label{figsic222}
\end{figure}

The difference of the former case can be seen as null, which is expected, since the original correction factor derived by equation (\ref{ch12}) should make the same correction irrespective of the reference frequency, as long as the spectral index is constant across the entire sky. The latter case does not have the same property, as the beam no longer integrates to unity but to different values at different frequencies after being \textit{stretched} by the ionospheric refraction. The difference, $\sim$ 0.2 K at lower frequencies is significant enough to overwhelm the global 21-cm signal. The residuals in Fig. \ref{figsic222} shows obvious changes in the residual when a different reference frequency is adopted. This is due to the intrinsic beam chromaticity being coupled with the chromatic ionospheric effects, that it no longer integrates to unity.

\begin{table}
	\centering
	\caption{The table shows the root mean square (RMS) of the residuals after fitting and the amplitude $|A_0|$, central frequency $\mu$, and width  $\sigma$ (standard deviation) of the fitted Gaussian signal. The reference (in red) is the original input signal used in the model. The residual increases with growing reference frequency. The overall fit shows an improvement as the reference frequency decreases, with the exception of width, which is shown to suffer greater reduction.}
	\label{tabrf}
	\begin{tabular}{rcccc} 
		
		&Residual& \multicolumn{3}{c}{Fitted Gaussian Signal} \\
		\hline
		 & RMS (K) & $|A_0|$ (K) & $\mu$ (MHz) & $\sigma$ (MHz)\\
		\hline
		\textcolor{BrickRed}{Reference} & -- & \textcolor{BrickRed}{0.53} &\textcolor{BrickRed}{78.1} &\textcolor{BrickRed}{18.7}\\
		90 MHz & $2.09 \times 10^{-1}$ & 0.87 & 80.8 & 16.7\\
		120 MHz & $2.96 \times 10^{-1}$ & 0.99 & 83.5 & 24.2\\
		150 MHz & $3.41 \times 10^{-1}$ & 0.99 & 86.1 & 25.6\\
		180 MHz & $3.59 \times 10^{-1}$ & 0.99 & 86.9 & 26.2\\

		\hline 
	\end{tabular}
\end{table}

\subsection{Different Gaussian Signals}

Apart from the signal motivated by \citet{edges}, two different global 21-cm signals of lower amplitudes ($\sim$ 0.1 K) and centred at other frequencies are also tested in the simulation. The parameters of these two signals lie within the possible range of high-redshift parameters suggested in \citet{cohen2017}. So far only two have been done; more cases are left as further work. Lower amplitudes are chosen because, as shown earlier, a signal of $\sim$ 0.5 K could be extracted, albeit with a significant residual, we want to see if a weaker signal would be completely overwhelmed. 

\begin{figure}
    \minipage{0.532\textwidth}
        \includegraphics[trim={0 0.7cm 0 6.48cm},clip,width=\linewidth]{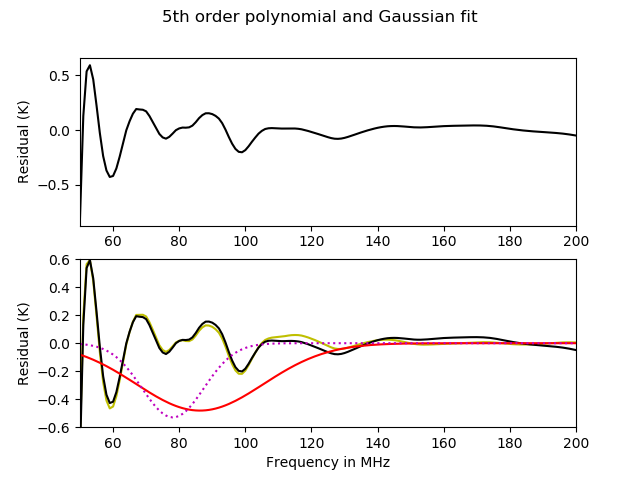}
    \endminipage\hfill \\
    \centering
    \minipage{0.532\textwidth}
        \includegraphics[trim={0 0.7cm 0 6.48cm},clip,width=\linewidth]{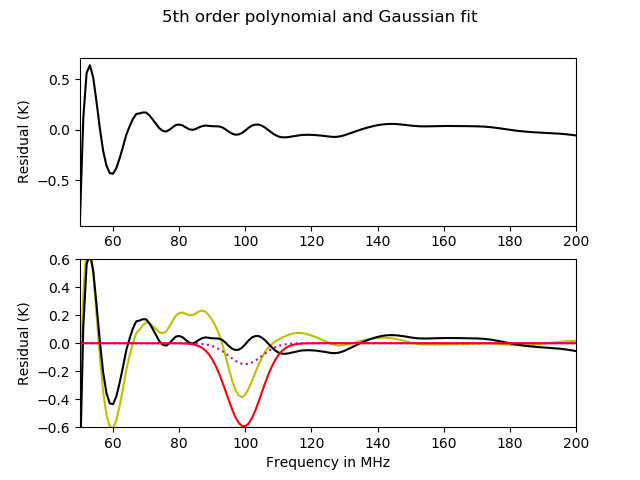}
    \endminipage\hfill \\
    \centering
    \minipage{0.532\textwidth}
        \includegraphics[trim={0 0 0 6.48cm},clip,width=\linewidth]{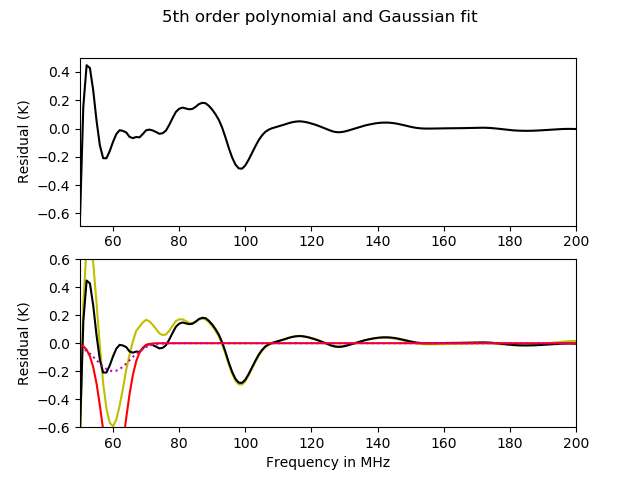}
    \endminipage\hfill
    \caption{The panels show the residuals of the signal modelled with ionospheric effects, fitted by equation (\ref{logpoly}) and (\ref{logpolygau}) using least-squares fit. The inserted Gaussian signals in the different panels:
    (top) $A_0 = 0.53$, $\mu = 78.1$ MHz, $\sigma=18.7$  MHz (middle) $A_0 = 0.15$, $\mu = 100$ MHz, $\sigma=10$ MHz (bottom)  $A_0 = 0.2$, $\mu = 60$ MHz, $\sigma=10$  MHz, integrated over 5 nights. Each panel consists of three different lines: \textbf{black} -- residual of the least-squares fit of a $5^{\text{th}}$ order log-polynomial and a Gaussian (equation \ref{logpolygau}), \textbf{olive} -- residual of the least-squares fit of a $5^{\text{th}}$ order log-polynomial only (equation \ref{logpoly}), \textbf{red} -- the Gaussian component of the combined least-squares fit (equation \ref{logpolygau}), and \textbf{dotted-magenta} -- the 21-cm signal added in the simulation.   The residuals can be as large as 0.6 K at the lower frequency range, similar to the case where the signal is larger, overwhelming the amplitude of the added signals. The fitted Gaussian signals overlap with the local minima in the D-layer residual pattern (Fig. \ref{figrandrannran}), and it is quite probable that the strong fitted Gaussian signal is only a compensation for the deviation from the pattern at the less stable points at the extrema.}
    \label{figrgausses}
\end{figure}

The results are shown in Fig. \ref{figrgausses}. The residuals of the fitted data generated in these simulations can be as large as 0.6 K at the lower frequency range, similar to the case where the signal is larger, overwhelming the amplitude of the added signals. What should be noted is that the fitted Gaussian signals shown in the plots apparently overlap with the local minima in the D-layer residual pattern mentioned previously, and it is quite probable that the strong fitted Gaussian signal is only a compensation for the deviation from the pattern at the less stable points, such as those at the extrema. The fit, therefore, should not be accepted as authentic. Moreover, the fitting used in this work is not sophisticated enough to properly extract the signal from a realistic foreground. Nevertheless, it is clear that the ionospheric effects have great potential to bury a weak global 21-cm signal of amplitude $\lessapprox 0.6$ K in an otherwise smooth foreground.

\subsection{REACH Pipeline Analysis}
\begin{figure}[H]
    \centering
    \minipage{0.42\textwidth}
        \includegraphics[width=\linewidth]{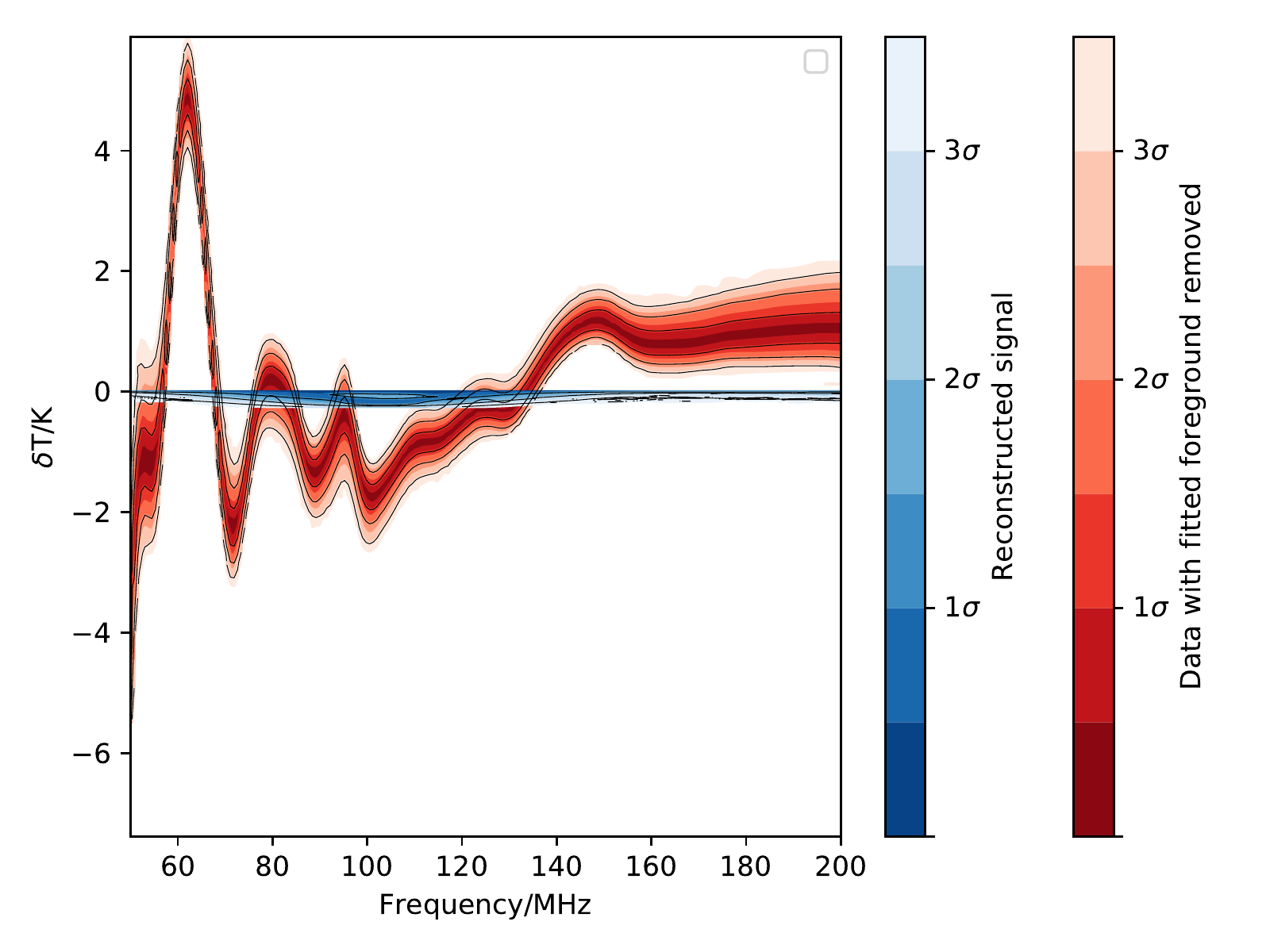}
    \endminipage\hfill
    \caption{Reconstructed signal (blue) and residual (red) of the data simulated with the ionosphere integrated every 15 minutes over 5 nights, 24 hours (the D-layer is constant in time and the F-layer varies with time). Parameters of the inserted Gaussian signal: $A_0 = 0.53$, $\mu = 78.1$ MHz, $\sigma=18.7$. The Gaussian signal cannot be recovered by the current version of the pipeline when the signal is contaminated by ionospheric effects.}
    \label{figrcs}
\end{figure}

We also tried to fit the simulated data using the current version of the REACH analysis pipeline, in which no feature to correct the ionospheric distortion has been added. In the pipeline, \textsc{PolyChord}, a state-of-the-art nested sampling algorithm for high-dimensional parameter spaces using Bayesian inference \citep{anstey}, is adopted. It tries to fit the data using the log likelihood given by
\begin{equation}
\begin{aligned}
\log(\mathcal{L}) = -\sum_\nu \frac{1}{2}\log{(2\pi\sigma^2)} + \frac{1}{2}\left(\frac{T_\mathrm{A}(\nu) -T_\mathrm{f}(\nu, \theta_{f})-T_{21}(\nu,\theta_{ \mathrm{sig} })}{\sigma} \right)^2,
\end{aligned}
\label{eqpolych}
\end{equation}
where $\sigma$ is a free parameter with a log uniform prior of 0.001-0.1 K. Fig. \ref{figrcs} shows the reconstructed Gaussian signal, in blue, and the residual, in red. It is apparent that the Gaussian signal cannot be recovered by the current version of the pipeline when the signal is contaminated by ionospheric effects, with a residual exceeding 4 K at lower frequencies, much higher than the intended signal detection at $\sim 0.5$ K.

\section{Conclusions}

The works shown in this paper can be summarised by the following points:

  (i) The total chromatic ionospheric effect is dominated by the D-layer absorption both in magnitude and chromatic distortion. The F-layer has a lower, yet significant, contribution. The residuals yielded by the simple log-polynomial fit seem to share a similar pattern, the local extrema occurring at the same frequencies in multiple cases when simulated without a signal.

 (ii) Irrespective of the extreme ionospheric conditions, foreground temperature can differ within a $\sim 2500$ K range, depending on the ionospheric condition. The relevant parameters include electron density, electron collision frequency, height and width in both layers.

 (iii) The global 21-cm signal modelled as a Gaussian can have its amplitude weakened and its shape distorted after being subjected to ionospheric interference based on the parameters used in this analysis.

  (iv) Preliminary results show that the chromaticity introduced by the ionosphere does not cancel out over time for a constant or a time-varying D-layer, with a time-varying F-layer, when using our simplified spatial model. It also suggests that the significance of the ionospheric effects greatly depend on the current ionospheric condition.

  (v) When simulated with a uniform spectral index, the beam chromaticity function and its modified counterpart have a limited use to correcting the antenna temperature subjected to chromatic ionospheric interference. Moreover, the modified equation yields different results when adopting different reference frequencies. The adopted chromaticity corrections have little to no use in the cases where spatially varying spectral indices are adopted.
 
 (vi) As precise removal of the foregrounds is critical in extracting the faint global 21-cm signal, the ionospheric model should to be included in data analysis pipelines to properly account for the chromaticity introduced by the ionosphere.  

The ionospheric model adopted in this work makes strong assumptions on the complex ionosphere, excluding spatial azimuthal variations and the dynamic phenomena that might affect the result significantly. Based on our results, we conclude that better ionospheric models are essential to understanding the chromatic global 21-cm signal distortion in ground-based experiments.

 A ionospheric model has not yet been included in the actual foreground model adopted in REACH. As a precise knowledge of the ionospheric condition is not currently achievable, the current ionospheric model can be improved by gaining further insights on the mathematical formulation of the ionospheric effects based on the observables as well as the simulations. With some final adjustments, it should eventually be deployed in the REACH analysis pipeline to enable a more accurate removal of the foregrounds.

\section*{Acknowledgements}
AF is supported by the Royal Society University Research Fellowship. WH was supported by a Gonville \& Caius research fellowship. EdLa and DA are supported by STFC.

\section*{Data Availability}
Two sets of simulation data (integrated antenna temperature) for reference are available at \doi{10.5281/zenodo.4105855}. The ionospheric data  collected from Lowell GIRO Data Center are available at \url{https://ulcar.uml.edu/DIDBase/}.



\bibliographystyle{mnras}
\bibliography{ref} 




\appendix

\section{Derivation of the Extra Chromaticity Correction Factor}
This is the full derivation of the extra chromaticity correction factor (equation \ref{sca} in section \ref{secda}):
\begin{equation}
\begin{aligned}
C_{\nu_2}(\nu) &= \frac{\int [T_\mathrm{f}(\nu_2,\Omega)- T_{\text{CMB}}]{B}(\nu,\Omega) d\Omega}
{\int [T_\mathrm{f}(\nu_2,\Omega)- T_{\text{CMB}}]{B}(\nu_2,\Omega)d\Omega} \\
&= \frac{\int [T_\mathrm{f}(\nu_1,\Omega) - T_{\text{CMB}}] (\nu_2/\nu_1)^{-\alpha} {B}(\nu,\Omega) d\Omega}
{\int [T_\mathrm{f}(\nu_1,\Omega) - T_{\text{CMB}}] (\nu_2/\nu_1)^{-\alpha} a(\Omega) {B}(\nu_1,\Omega)d\Omega} \\
& = \frac{(\nu_2/\nu_1)^{-\alpha} \int [T_\mathrm{f}(\nu_1,\Omega) - T_{\text{CMB}}] {B}(\nu,\Omega) d\Omega}
{(\nu_2/\nu_1)^{-\alpha} \int [T_\mathrm{f}(\nu_1,\Omega) - T_{\text{CMB}}] a(\Omega) {B}(\nu_1,\Omega)d\Omega}\\
& \approx \frac{\int T_\mathrm{f}(\nu_1,\Omega)  {B}(\nu,\Omega) d\Omega}{\int T_\mathrm{f}(\nu_1,\Omega) a(\Omega) {B}(\nu_1,\Omega)d\Omega}, \quad \text{when } T_{\text{CMB}} \text{ is small}\\
& = \frac{1}{A} \times \frac{\int T_\mathrm{f}(\nu_1,\Omega)  {B}(\nu,\Omega) d\Omega}{\int T_\mathrm{f}(\nu_1,\Omega) {B}(\nu_1,\Omega)d\Omega}  = \frac{C_{\nu_1}(\nu)}{A}.
\end{aligned}
\label{sca1}
\end{equation}


\bsp	
\label{lastpage}
\end{document}